\documentclass[letterpaper,12pt]{JHEP3}

\usepackage{amsmath}
\usepackage{amsfonts}

\def\e{\epsilon}
\def\G{\Gamma}

\def\Z{\mathcal{Z}}
\def\M{\mathcal{M}}

\def\O{\mathcal{O}}

\title{A Note on a standard embedding on half-flat manifolds}
\author{Tibra~Ali and Gerald~B. Cleaver\\ Physics Department\\ Baylor University\\ One Bear Place \# 97316\\ Waco, Texas 76798, U.S.A. \\E-mails: \email{Tibra\_Ali@baylor.edu}, \email{Gerald\_Cleaver@baylor.edu}}
\abstract{It is argued that the ten dimensional solution that corresponds to the compactification of $E_8\times E_8$ heterotic string theory on half-flat manifolds is the product space-time $\mathbb{R}^{1,2}\times \Z_7$ where $\Z_7$ is a generalized cylinder with $G_2$ holonomy. Standard embedding on $\Z_7$ then implies an embedding on the half-flat manifold which involves the torsionful connection rather than the Levi-Civita connection. This leads to the breakdown of $E_8\times E_8$ to $E_6\times E_8$, as in the case of the standard embedding on Calabi-Yau manifolds, which agrees with the result derived recently by Gurrieri, Lukas and Micu \cite{gurrieri3} using a different approach. Green-Schwarz anomaly cancellation is then implemented via the torsionful connection on half-flat manifolds.}
\begin{document}

%%%%%%%%%%%%%%%%%%%%%%%%%%%%%%%%%%%%%%%%%%
%%%%%%%%%%%%%%%%%%%%%%%%%%%%%%%%%%%%%%%%%%
\section{Introduction}
In recent years six-dimensional compact manifolds with $SU(3)$ structure have become a serious option for compactifications of string theories. We have been led to consider $SU(3)$-manifolds more general than Calabi-Yau (CY) manifolds both from theoretical and phenomenological considerations.

An $SU(3)$ manifold can be characterized by a geometric quantity known as the intrinsic torsion, which measures the deviation of the holonomy group of the Levi-Civita connection $\nabla^6$ from $SU(3)$ \cite{chiossi}. An interesting subclass of $SU(3)$-manifolds are \emph{half-flat} (HF) manifolds. On a HF manifold half of the possible components of the intrinsic torsion vanish (see \cite{chiossi} for more details). Certain HF manifolds, dubbed as \emph{HF mirror manifolds} in \cite{gurrieri2}, arise naturally from considerations of mirror symmetry in the presence of NS-NS background fluxes in type II string theories on CY manifolds \cite{gurrieri}. 

Although originating from type II theories these conjectured HF mirror manifolds are natural compactification spaces for the $E_8\times E_8$ heterotic string theory for two reasons. First, they are considered to be `small' deformations of CY's. Thus one expects the same spectrum of low-energy effective fields from them as one gets from CY compactifications. Indeed, as was noted in \cite{ali}, implicit in the work of \cite{gurrieri} is the fact that the Euler characteristic of a HF mirror manifold is the same as that of the `underlying' CY manifold, although the Betti numbers of the two manifolds are different.

The second reason for compactifying the heterotic string on HF mirror manifolds has to do with moduli-stabilization. Heterotic strings suffer from the apparent drawback that they offer fewer `fluxes' to be turned-on. On HF manifolds fluxes are geometrically encoded. In fact HF manifolds may be thought of as CY manifolds on which the Ricci-curvature has gained non-zero vacuum expectation value.\footnote{This point of view regarding the relationship between HF and CY manifolds is currently under investigation \cite{ali2}.} Thus to include some of the desirable features of flux compactifications \cite{giddings,kachru,grana,douglas} in the heterotic setting it is natural to turn to HF manifolds. 

The study of heterotic string theory on HF mirror manifolds was initiated in \cite{gurrieri2} and in a recent communication \cite{gurrieri3} the configuration of the background gauge field was explored. In \cite{gurrieri3} it has been noted that although the holonomy group of the Levi-Civita connection on HF manifolds is not $SU(3)$, a version of the standard embedding still leads to the breakdown of $E_8\times E_8$ to $E_6\times E_8$ as in the case of CY compactifications \cite{candelas}. In reaching this conclusion (which reverses a previous assertion  \cite{gurrieri2}, based on an analysis which excluded the gauge fields, that $E_8\times E_8$ breaks down to $SO(10)\times E_8$) the authors of \cite{gurrieri3} have made use of the `adiabatic principle' as well as the `Gukov superpotential' derived earlier in \cite{gurrieri2}. Very briefly, the adiabatic principle basically treats a HF mirror manifold like a CY, modulo the fact that some of the forms which used to be closed on the CY are no longer closed on the HF mirror manifold.  

Although, the adiabatic principle is very useful in deriving effective theories (e.g. see \cite{gurrieri, ali}) it still lacks complete mathematical justification.\footnote{For an important attempt in that direction, see \cite{kashani}.} Thus, in our opinion, it is extremely important to explore how far one can go in formulating important physical results in the heterotic string without the aid of such approximations. This is the subject of this note.

In CY compactification \cite{candelas} the modified Bianchi identity is solved by embedding the Levi-Civita spin connection into the gauge connection. This procedure has come to be known as the `standard embedding'. However, one of the central results used in this paper is the fact that the ten-dimensional low-energy effective action of the heterotic string is not unique \cite{hull}. Thus \emph{which} spin-connection is being embedded depends on the choice of low-energy variables. Thus, attaching the label `standard embedding' exclusively to embedding the Levi-Civita connection (which is just one of the possible spin-connections) would be misleading.

In contrast, the `non-standard' case is the one in which one solves the modified Bianchi identity without any embedding. In the CY case this involves solving the Donaldson-Yau-Ulhenbeck equation on holomorphic vector bundles on CY's.  (For an overview of this procedure see \cite{green}).

This is the usage followed in \cite{gurrieri3} and here, since we solve the modified Bianchi identity by embedding a spin-connection into the gauge connection, we use the term `standard embedding' to distinguish it from any approach that solves the Bianchi identity without an embedding technique.

In a strict sense, however, there is no such thing as \emph{the} standard embedding but an equivalence class of standard embeddings. This reflects the fact that due to an ambiguity in the anomaly there is an equivalence class of low-energy effective actions for the heterotic string theory (see \cite{hull} and our discussion below.) Thus our standard embedding is \emph{a} standard embedding. However, once the low-energy Lagrangian is chosen there should be a unique choice of standard embedding for a given compactification manifold. In what follows we shall always use the expression `standard embedding' in this sense.

The standard embedding on manifolds with $SU(3)$ structure is an important problem and, at least as far as the HF manifolds are concerned, it seems to us that there is a much more direct and transparent route to some of the same conclusions of \cite{gurrieri3}. %This is the subject of this note. 
Our approach in this note is thus complementary to the one taken in \cite{gurrieri3}. It is also an important justification of this note that our results are logically independent of any CY results (although, they are certainly inspired by them) and hence they hold true whether or not a certain HF manifold is thought of as adiabatic deformation of some CY. Our result is also independent of any arguments based on the Gukov superpotential.

Our conclusions are essentially the product of two results: Hitchin's theorem about the relationship between $G_2$-holonomy cylinders and HF manifolds \cite{hitchin}, and Hull's observations \cite{hull} regarding the nature of ambiguity in the Green-Schwarz anomaly cancellation condition.

In the next section we present our argument, relegating a technicality to an appendix.
%%%%%%%%%%%%%%%%%%%%%%%%%%%%%%%%%%
%%%%%%%%%%%%%%%%%%%%%%%%%%%%%%%%%%
\section{A Standard Embedding on Half-Flat Manifolds}
We start by asking the following question: What is the ten-dimensional space-time whose low-energy dynamics is described by the effective action of $E_8\times E_8$ heterotic string theory on HF manifolds?\footnote{In this paper the label `effective action' is used in two different contexts. First, there is the ten-dimensional effective action, which is the ten-dimensional supergravity action with the relevant anomaly cancellation terms. Secondly, there is the four-dimensional effective action which one obtains from dimensionally reducing a ten-dimensional effective action on a six-manifold with $SU(3)$ structure. It should be clear from the context which effective action we are referring to.} In the type II context supergravity no-go theorems forbid $\mathbb{R}^{1,3}$ as a solution to the effective theory. In fact the BPS solution turns out to be a domain wall \cite{mayer}. This solution can be taken over to the heterotic setting easily with the adjustment of the standard embedding described below. Thus it is clear that the ten-dimensional solution that is relevant for the case at hand is the \emph{lift} of the domain wall which is the direct product space-time $\mathbb{R}^{1,2}\times \Z_7$ where $\Z_7$ is a $G_2$-holonomy generalized cylinder\footnote{The term `generalized cylinder' was introduced in \cite{bar} to describe space-times whose metric has the form of eq.(\ref{eq:cylinder}).} with the natural metric
\begin{equation}
\begin{split}
ds^2 &= g_{MN} dx^M dx^N\\
&= dz^2 + g(z,y)_{mn} dy^m dy^n \label{eq:cylinder}
\end{split}
\end{equation}
where $x^M$ with $M=1, \dots, 6, z$ are coordinates on $\Z_7$ and $y^m$ with $m=1, \dots, 6$ are coordinates on six dimensional hypersurfaces ${}^z\M_6$ with the metric $g(z,y)_{mn}$ for a fixed value of $z$. Then according to Hitchin \cite{hitchin} ${}^z\M_6$ are HF manifolds.

%In a supersymmetric theory, with bosonic and ferminonic fields schematically denoted by $B$ and $F$, respectively, the supersymmetry transformations are of the form

%\begin{equation}\begin{split}
%\delta_\epsilon B &= \epsilon\left\{ Q, F\right\} \\
%\delta_\epsilon F &= \epsilon \left[ Q, B\right] \label{eq:susy} \end{split}
%\end{equation}
%where $\epsilon$ represents the Grassmann supersymmetry parameter and $Q$ the supercharges. A purely bosonic background (some of $B$'s nonzero and $F=0$) is then considered supersymmetric if $\delta_\epsilon B= \delta_\epsilon F =0$ for this background. The first set of these equations is automatically satisfied by virture of $F=0$. Thus, for a bosonic background to be  one needs to check that the second set of equations are also satisfied. Here we are interested in supersymmetric solutions to the theory because we want the final four dimensional theory to have $\mathcal{N}=1$ supersymmetry. One can look for solutions which break supersymmetry at the perturbative level, but then quantum corrections to the low energy theory make any prediction invalid. In general, one wants to invoke non-perturbative effects to break supersymmetry. In this note we shall not be concerned with supersymmetry breaking mechanisms.

We now verify that the above metric ansatz satisfy the supersymmetry conditions. The supersymmetry conditions, in the variables of \cite{candelas,green}, are:

\begin{equation}\begin{split}
\delta \psi_{\hat{M}} &= \nabla_{\hat{M}} \eta + \frac{1}{32\phi}\left(\G_{\hat{M}}{}^{\hat{P}\hat{Q}\hat{R}} - 9 \delta_{\hat{M}}^{\hat{P}} \G^{\hat{Q}\hat{R}}\right) H_{\hat{P}\hat{Q}\hat{R}} \e\\
&=0\\
\delta\lambda &= \frac{1}{\sqrt{2}\phi} \left(-\G^{\hat{M}} \partial_{\hat{M}} \phi+\frac{1}{8} \G^{\hat{P}\hat{Q}\hat{R}}H_{\hat{P}\hat{Q}\hat{R}}\right) \e=0\\
\delta \chi&=-\frac{1}{4\sqrt{\phi}} \G^{\hat{P}\hat{Q}} F_{\hat{P}\hat{Q}} \e=0 \label{eq:susy}
\end{split}\end{equation}
where the hatted Latin indices cover all ten dimensions of space-time. $\nabla_{\hat{M}}$ is the ten-dimensional Levi-Civita connection. $\psi_{\hat{M}}$, $\lambda$ and $\chi$ denote the gravitino, dilatino and gaugino fields, respectively. $\e$ is the local supersymmetry parameter. $F_{\hat{P}\hat{Q}}$ is the Yang-Mills field with gauge indices suppressed, $\phi$ is the dilaton and $H_{\hat{P}\hat{Q}\hat{R}}$ is the three-form gauge field given by:
\begin{equation}\begin{split}
H = dB + \O (\omega)- \O (A)
\end{split}\end{equation}
with $B$, $\O(\omega)$ and $\O(A)$ as the two-form potential, Chern-Simons 3-forms in the Lorentz and the gauge sectors, respectively. The $\G$'s above are the antisymmetrized Dirac matrices (i.e., Clifford algebra elements.) We are also using units in which the gravitational and Yang-Mills coupling constants are chosen to be unity.

The second equation of (\ref{eq:susy}) is satisfied by choosing an ansatz $d\phi=H=0$. With this choice in the $\mathbb{R}^{1,2}\times \Z_7$ background it is easy to see that the gravitino variation condition (the first equation of  (\ref{eq:susy})) is satisfied. On $\Z_7$ the Majorana spinor $\epsilon$ satisfies
\begin{equation}
\begin{split}
\nabla^7_M \epsilon =0 \label{eq:g2spinor}
\end{split}
\end{equation}
where $\nabla^7$ is the Levi-Civita connection on $\Z_7$. The integrability condition for the above equation is
\begin{equation}
\begin{split}
R^7_{MNPQ}\Gamma^{PQ} \epsilon =0. \label{eq:g2integrability}
\end{split}
\end{equation}
where $\Gamma^{PQ}$ is the antisymmetrized product of two Dirac matrices on $\Z_7$. These equations imply that the Riemannian holonomy group of $\Z_7$ is contained in $G_2$. To solve the gaugino supersymmetry condition we adopt 
\begin{equation}
\begin{split}
A = \omega^7 \label{eq:g2standard}
\end{split}
\end{equation}
where $A$ is the $E_8\times E_8$ Yang-Mills gauge potential and $\omega^7$ is the spin-connection with $G_2$ holonomy that corresponds to the Levi-Civita connection $\nabla^7$. This is the standard embedding for $G_2$ holonomy background. Because of (\ref{eq:g2integrability}) and (\ref{eq:g2standard}) we then have
\begin{equation}\begin{split}
F_{MN} \Gamma^{MN}\epsilon =0.
\end{split}\end{equation}
Thus we see that the remaining condition, the last equation of (\ref{eq:susy}), is satisfied. 
 
The choice eq.(\ref{eq:g2standard}) then leads to the Green-Schwarz anomaly condition
\begin{equation}
dH = \mathrm{tr} R \wedge R - \frac{1}{30} \mathrm{Tr} F\wedge F \label{eq:anomaly}
\end{equation}
being satisfied with $H=0$. 

What we have outlined above is a special case of an `instanton' solution explored in more detail in \cite{ivanov}.
If we were interested in the three dimensional effective action of the compactification on a (say, compact) $G_2$ holonomy manifold then the above construction would lead $E_8\times E_8$ to break down to $F_4\times E_8$. This is because the commutant of $G_2$ in $ E_8$ is $F_4$. This would be consistent with the fact that there is no notion of chirality in three dimensions (since $F_4$ doesn't lead to chiral multiplets). However, we are interested in knowing what is the condition that \emph{descends} on the HF slices ${}^z\M_6$ from the condition (\ref{eq:g2standard}).

Since HF manifolds are manifolds with $SU(3)$ structure there must exist an almost complex structure $J_m{}^n$ and a complex three-form $\Omega_{pqr}$ which is of type (3,0) with respect to the almost complex structure. These quantities are globally defined, or in other words there must be a connection $\widetilde{\nabla}^6$ with respect to which these quantities are covariantly constant:
\begin{equation}\begin{split}
\widetilde{\nabla}^6_m J_p{}^q &=0 \\
\widetilde{\nabla}^6_m \Omega_{pqr} &=0. \label{eq:su3struc}
\end{split}\end{equation}
This is equivalent to the statement that there exists a globally defined Majorana spinor $\e'$ which is also covariantly constant
\begin{equation}\begin{split}
\widetilde{\nabla}^6_m \epsilon' =0. \label{eq:hfspinor}
\end{split}\end{equation}
$J$ and $\Omega$ can be expressed as bilinears of $\e'$ (see our appendix for more details). However the connection appearing in above is not the Levi-Civita connection $\nabla^6$. It differs from $\nabla^6$ by the intrinsic torsion whose detailed form is given in the appendix. The integrability of the condition of (\ref{eq:hfspinor}) is given by
\begin{equation}\begin{split}
\widetilde{R}_{mnpq}^6 \Gamma^{pq}\epsilon' =0. \label{eq:hfintegrability}
\end{split}\end{equation}
In the above equation $\widetilde{R}_{mnpq}^6$ is the curvature of the connection $\widetilde{\nabla}^6$. Note that the torsion term drops out due to (\ref{eq:hfspinor}). The globally defined spinor $\epsilon'$ that defines the $SU(3)$ structure is the same spinor whose constancy in seven-dimensions leads to $G_2$ holonomy\footnote{In principle, $\e'$ has `space-time' components as well but we shall not concern ourselves with the product structure of $\e'$ in terms of spinors in $\mathbb{R}^{1,2}$ and $\Z_7$.}, i.e.
\begin{equation}\begin{split}
\epsilon' =\epsilon. \label{eq:equality}
\end{split}\end{equation}
The above equality is, of course, guaranteed by Hitchin's theorem \cite{hitchin}. In the appendix we give an explicit demonstration of the fact that (\ref{eq:hfspinor}) indeed follows from (\ref{eq:g2spinor}) using the expressions developed in \cite{ali} which relates the intrinsic torsion of ${}^z\M_6$ and the extrinsic curvature of the embedding of ${}^z\M_6$ in $\Z_7$.

Let us now see what is the condition that descends from (\ref{eq:g2standard}) on the six dimensional half-flat slices. To see this we rewrite the $m=1,\dots, 6$ \emph{components} of (\ref{eq:g2standard}):
\begin{equation}
A_m = \omega^7_m. \label{eq:standard}
\end{equation}
In writing the above relation we have excluded $m=z$ component of (\ref{eq:g2standard}). We have also suppressed the  tangent-space and gauge indices. 

For the truncated condition (\ref{eq:standard}) to make sense as a six dimensional equation the right hand side must have an interpretation in six dimensions. In fact it does: it is simply the torsionful spin-connection whose holonomy is $SU(3)$. Thus we conclude that the seven dimensional standard embedding implies the following condition on the HF slices:
\begin{equation}\begin{split}
A = \widetilde{\omega}^6 \label{eq:hfstandard}
\end{split}\end{equation}
where $\widetilde{\omega}^6$ is the metric-compatible \emph{torsionful} spin-connection whose holonomy is $SU(3)$. Note that this is \emph{not} a Riemannian holonomy group.

Some time ago, it was pointed out by Hull \cite{hull} that the anomaly in a gauge theory is always ambiguous up to a change of the connection by a tensor quantity. Changing to a new connection in the action (and hence the path-integral) simply implies a corresponding change in the counter-term that is needed to cancel the anomaly. 

With the condition chosen in (\ref{eq:hfstandard}) the Green-Schwarz anomaly cancellation condition is then no longer (\ref{eq:anomaly}) but is instead given by
\begin{equation}
d\widetilde{H} = \mathrm{tr} \widetilde{R}^6 \wedge \widetilde{R}^6 - \frac{1}{30} \mathrm{Tr} F\wedge F \label{eq:anomaly2}
\end{equation}
with $\widetilde{R}^6$ being the same curvature that appeared in (\ref{eq:hfintegrability}). $\widetilde{H}$ in the above equation is defined by
\begin{equation}
\widetilde{H} = dB + \O (\widetilde{\omega}^6) - \O (A)
\end{equation}
The `new' anomaly cancellation condition (\ref{eq:anomaly2}) is then solved by $\widetilde{H}=0$ and (\ref{eq:hfstandard}). This implies that the ten-dimensional low-energy effective theory natural for half-flat manifolds is a theory that is \emph{different} from the ones that are usually considered in the literature. We refer to Hull's original paper \cite{hull} for more details about the steps in obtaining this `new' effective action. We hope to present this effective action in a future publication \cite{ali2}.

This leads us to the main conclusion of this paper:

\medskip 

\noindent \emph{Equation (\ref{eq:hfstandard}) is then a valid standard embedding on HF manifolds. Since the holonomy of $\widetilde{\omega}^6$ is $SU(3)$, eq.(\ref{eq:hfstandard}) implies $E_8\times E_8$ breaks down to $E_6\times E_8$ just as in the case of CY compactifications.}
%%%%%%%%%%%%%%%%%%%%%%%%%%%%%%%%%%%%%%%%%%
%%%%%%%%%%%%%%%%%%%%%%%%%%%%%%%%%%%%%%%%%%
\section{Conclusions and Outlook}
In this paper we have argued that the ten dimensional solution that corresponds to the compactification of $E_8\times E_8$ heterotic string theory on a half-flat manifold is the direct product space-time $\mathbb{R}^{1,2}\times \Z_7$ where $\Z_7$ is a generalized cylinder with $G_2$ holonomy \emph{\`a la} Hitchin \cite{hitchin}. The supersymmetry conditions and the Green-Schwarz anomaly cancellation condition are then satisfied by embedding the Levi-Civita connection of $\Z_7$ in the gauge connection. This implies a standard embedding on the half-flat slices which is given by (\ref{eq:hfstandard}) which leads to the breakdown of $E_8\times E_8$ to $E_6\times E_8$. However, this implies that the natural variables for HF manifolds are not any of the standard formulations of the ten-dimensional low-energy effective actions of heterotic string theory. But the existence of such a formulation is guaranteed by an ambiguity in the Green-Schwarz anomaly \cite{hull}. It is important to note that, unlike \cite{gurrieri3}, we did not assume that the HF manifolds had to be some sort of `small' or `adiabatic' deformation of an underlying CY. Once the ansatz (\ref{eq:hfstandard}) is adopted the anomaly cancellation condition is satisfied \emph{exactly} up to $\mathcal{O}(\alpha')$. Our results are valid for \emph{any} half-flat manifold (including nilmanifolds). However, for phenomenological purposes one is mainly interested in HF mirror manifolds. It is our hope that the fact the approach taken here is independent of some of the approximations made in \cite{gurrieri3} puts their effort on firmer ground.

Since the full ten dimensional solution breaks down to $F_4\times E_8$, it is a prediction of our work that the domain wall solutions of the effective theory on HF manifolds will spontaneously break $E_6$ down to $F_4$.  To see this more clearly recall that in the type II setting the BPS `ground-state' of the low energy effective theory are domain walls \cite{mayer}. We expect the same to be true in the heterotic string. Since the four dimensional solution  must lift up to $\mathbb{R}^{1,2}\times\Z_7$ with the gauge  fields given by (\ref{eq:g2standard})  it follows
that the ground-state will break $E_6$ down to $F_4$. How this is actually implemented in the four-dimensional effective action will be presented in a following paper \cite{ali2}. This prediction is in contrast with that of \cite{gurrieri3} where they argue that $E_6$ should spontaneously break down to $SO(10)$. In passing we note that the decomposition of $E_8$ in terms of $F_4$ and $ G_2$ which comes out of HF manifolds is reminiscent of the group-structure of the unification scenario recently proposed by Lisi \cite{lisi}.

%To see this more clearly recall that in the type II setting the BPS `ground-state' of the low energy effective theory are domain walls \cite{mayer}. Since this must lift up to $\mathbb{R}^{1,2}\times \Z_7$ with the gauge fields given by \ref{eq:g2standard} it follows that the ground-state will break $E_6$ down to $F_4$. How this is implemented in the four-dimensional effective action will be presented in a following paper \cite{ali2}. 

Because our conclusions are based on a \emph{bona fide} ten-dimensional solution one cannot \emph{a priori} comment on the standard embeddings on $SU(3)$ manifolds more general than HF manifolds. In our view one needs to first look at the ten-dimensional solutions before making ans\"atze for dimensional reductions on six dimensional manifolds with an arbitrary $SU(3)$-structure. In light of the swamp-land conjecture \cite{vafa} it is important, in our opinion, to keep in the background the full ten-dimensional solution.

Since the ten-dimensional effective actions for heterotic string theories contain higher order curvature terms it is not immediately clear that, despite our standard embedding, all the curvature dependent terms are computable on HF mirror manifolds. However, since the Ricci-curvature for HF manifolds have now been computed in terms of variables familiar from CY compactifications \cite{ali} it seems not unlikely that one can compute the effective action in these new variables. This effective action would be a consistency check of the one derived in \cite{gurrieri3} and would perhaps be related to the latter action by a change of variables. This and other issues are now under investigation \cite{ali2}. 
%%%%%%%%%%%%%%%%%%%%%%%%%%%%%%%%%%%%%%%%%%
%%%%%%%%%%%%%%%%%%%%%%%%%%%%%%%%%%%%%%%%%%
\section*{Note Added}
After our paper appeared on the archives, it was brought to our attention that similar results were presented in \cite{prezas} for a sub-class of half-flat manifolds known as nearly K\"ahler manifolds.
%%%%%%%%%%%%%%%%%%%%%%%%%%%%%%%%%%%%%%%%%%%
%%%%%%%%%%%%%%%%%%%%%%%%%%%%%%%%%%%%%%%%%%%
\acknowledgments
It is a pleasure to thank S.\ Majhi for insightful and stimulating discussions. We would also like to thank Andrei Micu for his comments on the first version of this paper.
%%%%%%%%%%%%%%%%%%%%%%%%%%%%%%%%%%%%%%%%%%%%%%%%%%%%%%%%%%%%%%%%%%%%%%%%%%%%%%%%%%%%%%%%%%%%%%%%%%%%%%%%%%%%%%%
\appendix
\section{Appendix: $SU(3)$ holonomy and $G_2$ holonomy}
We are interested in the six-dimensional condition that descends from the seven-dimensional standard embedding that we discussed above. Hitchin's theorem implies that the relevant connection in six dimensions is the one with torsion. In this section we verify that explicitly. For more details on the origin of the following formulae the reader is referred to \cite{ali}.

An $SU(3)$ manifold admits a globally defined two form $J$ (related to an almost complex structure $J_m{}^n$) and a complex three form $\Omega=\Omega^+ + i \Omega^-$ which is of type $(3,0)$ with respect to $J_m{}^n$. On a Calabi-Yau manifold both of these forms are closed. On a half-flat manifold one has instead
\begin{equation}\begin{split}
J\wedge dJ &=0\\
d\Omega^- &=0
\end{split}\end{equation}
To show that (\ref{eq:hfspinor}) follows from (\ref{eq:g2spinor}):
\renewcommand{\theequation}{\arabic{section}.\arabic{equation}}
\setcounter{section}{2}
\setcounter{equation}{3}
\begin{equation}
\begin{split}
\nabla^7_M \epsilon =0 
\end{split}
\end{equation}
we start by looking at the components of this latter equation lying along the half-flat slices:
\renewcommand{\theequation}{\Alph{section}.\arabic{equation}}
\setcounter{section}{1}
\setcounter{equation}{1}
\begin{equation}
\nabla^7_m \epsilon =0.
\end{equation}
Since $m=1,\dots, 6$, the above equation is not obviously tensorial in six dimensions. 

Let us denote by $\nabla^6$ the Levi-Civita connection in six-dimensions. The Gau\ss -Weingarten equation then gives \cite{ali}:
\begin{equation}
\nabla^7_m \epsilon =\nabla^6_m \epsilon + \frac{1}{2} K_m^n \Gamma_n \Gamma^7 \epsilon.  \label{eq:gausswein}
\end{equation}     
Where $K_{mn}$ is the second fundamental form of the embedding of the half-flat manifold in a $G_2$ holonomy cylinder.
In \cite{ali} it was shown that for half-flat manifolds the intrinsic contorsion $\kappa_{rst}$ (which is equivalent to the intrinsic torsion) is related to $K_{mn}$ via
\begin{equation}
\kappa_{rst} = \frac{1}{2} K_r^m \Omega^+_{mst}
\end{equation}
and its inverse
\begin{equation}
K_r^p = \frac{1}{2} \Omega^+{}^{pst} \kappa_{rst}.
\end{equation}
where $\Omega^+$ is the real part of the three form $\Omega$. Then (\ref{eq:gausswein}) becomes
\begin{equation}
\nabla^7_m \epsilon = \nabla^6_m \epsilon + \frac{1}{4} \Omega^+{}^{nst}\kappa_{mst} \Gamma_n \Gamma^7\epsilon.
\end{equation}
We shall now show the connection on the right hand side (and hence the truncated connection on the left hand side) has $SU(3)$ holonomy. Let us first recall the following Clifford algebra identities:
\begin{equation}
\begin{split}
\Gamma_{mnp} &= \Gamma_{mn} \Gamma_p + g_{pm} \Gamma_n - g_{pn} \Gamma_m \\
 &= \Gamma_{m} \Gamma_{np} + g_{pm} \Gamma_n - g_{mn} \Gamma_p 
\end{split}
\end{equation}
We also need the following Fierz identity, which is simply a statement of the decomposition of identity on spinorial vector space in six spatial dimensions:
\begin{equation}
{\mathbf 1}_{8\times 8} = \epsilon \bar{\epsilon} + \Gamma_7 \epsilon \bar{\epsilon} \Gamma^7 +\Gamma_m \epsilon \bar{\epsilon} \Gamma^m
\end{equation}
We can now consider
\begin{equation}\begin{split}
\Gamma_{mn}\Gamma^7 \epsilon &= \left\{\epsilon \bar{\epsilon} + \Gamma_7 \epsilon \bar{\epsilon} \Gamma^7 +\Gamma_r \epsilon \bar{\epsilon} \Gamma^r \right\}\Gamma_{mn}\Gamma^7 \epsilon\\
&= \epsilon \bar{\epsilon} \Gamma_{mn}\Gamma^7\epsilon + \Gamma_r \epsilon \bar{\epsilon} \Gamma^r \Gamma_{mn} \Gamma^7 \epsilon
\end{split}
\end{equation}
where in going from the first line to the second we have used the fact that for any commuting Majorana spinor $\epsilon$ in six dimensions we have
\begin{equation}
\bar{\epsilon} \Gamma_7 \Gamma_{mn} \Gamma^7 \epsilon =\bar{\epsilon} \Gamma_{mn} \epsilon=0 .
\end{equation}
Next we use the following definitions (defined in \cite{ali})
\begin{equation}\begin{split}
J_{mn} &= i \bar{\epsilon} \Gamma_7 \Gamma_{mn} \epsilon \\
\Omega^+_{mnp} &= \bar{\epsilon} \Gamma_{mnp} \Gamma_7 \epsilon
\end{split}
\end{equation}
in the above identity and obtain
\begin{equation}
\Gamma_{mn} \epsilon = - i J_{mn}\Gamma^7 \epsilon - \Omega^+_{rmn} \Gamma^r \Gamma^7 \epsilon. 
\end{equation}
Using this in (\ref{eq:gausswein}) we get
\begin{equation}\begin{split}
\nabla^7_m \epsilon = \nabla^6_m \epsilon + \frac{1}{4} \kappa_{mst} \left\{ - i J^{st} \Gamma^7 - \Gamma^{st}\right\} \epsilon. \label{eq:gauss2}
\end{split}\end{equation}
In \cite{chiossi} it was shown that the torsion or the contorsion on an $SU(3)$ manifold decomposes into five $SU(3)$ modules. On a half-flat manifold some of these modules survive which we denote by $W^+_1$, $W^+_2$ and $W_3$. More details on these modules relevant for the present computation can be found in \cite{ali}. Here we note that $W^+_1$ is a real scalar, $W^+_2$ is a real, primitive (1,1)-type 2-form and $W_3$ is a complex, primitive (2,1)-type 3-form.
Now it is clear from the following expression of the contorsion tensor on a half-flat manifold \cite{ali}
\begin{equation}\begin{split}
\kappa_{srt} &= \frac{1}{4} W^+_1 \Omega^+_{srt} + \frac{1}{4} (W^+_2)_{sq} \Omega^q{}_{rt} \\
&+ \frac{i}{2} \left\{ (W_3)_{suv} \Pi^+_r {}^u \Pi^+_t {}^v - (\overline{W}_3)_{suv} \Pi^-_r {}^u \Pi^-_t {}^v \right\}
\end{split}
\end{equation}
that
\begin{equation}
J^{rt} \kappa_{srt} =0.
\end{equation}
Thus (\ref{eq:gauss2}) becomes
\begin{equation}\begin{split}
\nabla^7_m \epsilon &= \nabla^6_m \epsilon - \frac{1}{4} \kappa_{mst} \Gamma^{st} \epsilon \\
&\equiv \widetilde{\nabla}^6_m \epsilon.
\end{split}
\end{equation}
As a consequence of (\ref{eq:g2spinor}) we have for the connection $\widetilde{\nabla}$ with contorsion $\kappa$
\begin{equation}
\widetilde{\nabla}^6_m \epsilon =0
\end{equation}
which means that the holonomy of $\widetilde{\nabla}^6$ is $SU(3)$. Hence $\epsilon=\epsilon'$ where the latter is the globally defined spinor on the HF manifold which determines its $SU(3)$ structure.

\end{document}